\documentstyle[aps,manuscript,epsf]{revtex} 
\def\baselinestretch{1.25}
\clearpage

\begin{document}

\input{psfig}
\begin{flushright}
CDF/PUB/BOTTOM/PUBLIC/4966\\
FERMILAB-Pub-99/210-E \\
\today \\
\end{flushright}
\vspace{0.20in}
\begin {center}
\begin {Large}
{\bf \boldmath
Measurement of the 
$B^0 {\overline B}$$^0$      
oscillation frequency \\
using 
$\ell^- D^{*+}$ pairs and 
lepton
flavor
tags
}
\footnote{Submitted to Physical Review D. }
\end {Large}
\\
\end{center}

\font\eightit=cmti8
\def\r#1{\ignorespaces $^{#1}$}
\hfilneg
\begin{sloppypar}
\noindent
T.~Affolder,\r {21} H.~Akimoto,\r {42}
A.~Akopian,\r {35} M.~G.~Albrow,\r {10} P.~Amaral,\r 7 S.~R.~Amendolia,\r {31} 
D.~Amidei,\r {24} J.~Antos,\r 1 
G.~Apollinari,\r {35} T.~Arisawa,\r {42}
% ...
\end{sloppypar}
\vskip .026in
\begin{center}
(CDF Collaboration)
\end{center}

\vskip .026in
\begin{center}
\r 1  {\eightit Institute of Physics, Academia Sinica, Taipei, Taiwan 11529, 
Republic of China} \\
\r 2  {\eightit Argonne National Laboratory, Argonne, Illinois 60439} \\
\r 3  {\eightit Istituto Nazionale di Fisica Nucleare, University of Bologna,
I-40127 Bologna, Italy} \\
\r 4  {\eightit Brandeis University, Waltham, Massachusetts 02254} \\
\r 5  {\eightit University of California at Los Angeles, Los 
Angeles, California  90024} \\  
\r 6  {\eightit Instituto de Fisica de Cantabria, University of Cantabria, 
39005 Santander, Spain} \\
\r 7  {\eightit Enrico Fermi Institute, University of Chicago, Chicago, 
Illinois 60637} \\
\r 8  {\eightit Joint Institute for Nuclear Research, RU-141980 Dubna, Russia}
\\
\r 9  {\eightit Duke University, Durham, North Carolina  27708} \\
\r {10}  {\eightit Fermi National Accelerator Laboratory, Batavia, Illinois 
60510} \\
\r {11} {\eightit University of Florida, Gainesville, Florida  32611} \\
\r {12} {\eightit Laboratori Nazionali di Frascati, Istituto Nazionale di Fisica
               Nucleare, I-00044 Frascati, Italy} \\
\r {13} {\eightit University of Geneva, CH-1211 Geneva 4, Switzerland} \\
\r {14} {\eightit Harvard University, Cambridge, Massachusetts 02138} \\
\r {15} {\eightit Hiroshima University, Higashi-Hiroshima 724, Japan} \\
\r {16} {\eightit University of Illinois, Urbana, Illinois 61801} \\
\r {17} {\eightit The Johns Hopkins University, Baltimore, Maryland 21218} \\
\r {18} {\eightit Institut f\"{u}r Experimentelle Kernphysik, 
Universit\"{a}t Karlsruhe, 76128 Karlsruhe, Germany} \\
\r {19} {\eightit Korean Hadron Collider Laboratory: Kyungpook National
University, Taegu 702-701; Seoul National University, Seoul 151-742; and
SungKyunKwan University, Suwon 440-746; Korea} \\
\r {20} {\eightit High Energy Accelerator Research Organization (KEK), Tsukuba, 
Ibaraki 305, Japan} \\
\r {21} {\eightit Ernest Orlando Lawrence Berkeley National Laboratory, 
Berkeley, California 94720} \\
\r {22} {\eightit Massachusetts Institute of Technology, Cambridge,
Massachusetts  02139} \\   
\r {23} {\eightit Institute of Particle Physics: McGill University, Montreal 
H3A 2T8; and University of Toronto, Toronto M5S 1A7; Canada} \\
\r {24} {\eightit University of Michigan, Ann Arbor, Michigan 48109} \\
\r {25} {\eightit Michigan State University, East Lansing, Michigan  48824} \\
\r {26} {\eightit University of New Mexico, Albuquerque, New Mexico 87131} \\
\r {27} {\eightit The Ohio State University, Columbus, Ohio  43210} \\
\r {28} {\eightit Osaka City University, Osaka 588, Japan} \\
\r {29} {\eightit Universita di Padova, Istituto Nazionale di Fisica 
          Nucleare, Sezione di Padova, I-35131 Padova, Italy} \\
\r {30} {\eightit University of Pennsylvania, Philadelphia, 
        Pennsylvania 19104} \\   
\r {31} {\eightit Istituto Nazionale di Fisica Nucleare, University and Scuola
               Normale Superiore of Pisa, I-56100 Pisa, Italy} \\
\r {32} {\eightit University of Pittsburgh, Pittsburgh, Pennsylvania 15260} \\
\r {33} {\eightit Purdue University, West Lafayette, Indiana 47907} \\
\r {34} {\eightit University of Rochester, Rochester, New York 14627} \\
\r {35} {\eightit Rockefeller University, New York, New York 10021} \\
\r {36} {\eightit Rutgers University, Piscataway, New Jersey 08855} \\
\r {37} {\eightit Texas A\&M University, College Station, Texas 77843} \\
\r {38} {\eightit Texas Tech University, Lubbock, Texas 79409} \\
\r {39} {\eightit Istituto Nazionale di Fisica Nucleare, University of Trieste/
Udine, Italy} \\
\r {40} {\eightit University of Tsukuba, Tsukuba, Ibaraki 305, Japan} \\
\r {41} {\eightit Tufts University, Medford, Massachusetts 02155} \\
\r {42} {\eightit Waseda University, Tokyo 169, Japan} \\
\r {43} {\eightit University of Wisconsin, Madison, Wisconsin 53706} \\
\r {44} {\eightit Yale University, New Haven, Connecticut 06520} \\
\end{center}


\newpage
\vspace{0.25in}
\begin{abstract}
The oscillation frequency 
$\Delta m_d$  
of 
$B^0 \overline{B}$$^0$ mixing 
is measured using the
partially reconstructed
semileptonic decay 
$\overline{B}$$^0 \rightarrow \ell^- {\overline \nu} D^{*+} X$.
The data sample was collected with the CDF detector 
at the Fermilab Tevatron collider
during 1992--1995 
by triggering on 
the 
existence of two lepton candidates in an event, 
and corresponds to 
about 110~pb$^{-1}$ of
$\bar{p} p$ collisions at $\sqrt{s} =$ 1.8 TeV.
We estimate the proper decay time of the $\overline{B}$$^0$ meson
from 
the 
measured decay length and reconstructed momentum
of the $\ell^- D^{*+}$ system. The charge of the lepton
in the final state identifies the flavor of the 
${\overline B}$$^0$  meson at its decay.
The second lepton in the event 
is used 
to infer the flavor 
of the ${\overline B}$$^0$  meson
at production. 
We measure the oscillation frequency to be
$\Delta m_d = 
0.516 \pm 0.099 \;
^{\, + \, 0.029} _{\, - \, 0.035 }$~ps$^{-1}$,
where the first uncertainty is statistical and the second is systematic.
\end{abstract}
\vspace{1cm}
{PACS numbers: 14.40.Nd, 13.20.He} \vspace{0.2in} \\
%

\newpage
\section{Introduction}
Particle-antiparticle mixing in
the $B^0 \overline{B}$$^0$ system 
has been known for a decade now~\cite{Argus_mix}. 
The phenomenon can be understood as a second-order weak interaction
effect. 
The frequency of the oscillation between the two states
corresponds to 
the mass difference $\Delta m_d$
between the two mass eigenstates
of the $B^0 \overline{B}$$^0$ system, $B^0_H$ and $B^0_L$.
It can be calculated~\cite{Lim}
from 
box diagrams,
where contributions of 
the top quark in the loop
are dominant.
Measurements of 
$B^0 \overline{B}$$^0$ mixing
can 
therefore 
determine   
the magnitude 
of the 
Kobayashi-Maskawa matrix~\cite{CKM} 
element $V_{td}$.

Experiments at the $\Upsilon(4S)$ resonance 
have measured 
the probability of 
mixing, $\chi_d$,
integrated over decay time~\cite{chi_d}.
Experiments 
at 
LEP~\cite{new_LEP,new_LEP_common} 
and 
the Collider Detector at Fermilab (CDF)~\cite{CDF_mix},
where $B$ hadrons are produced at higher energies,
examine the time development of 
mixing 
and measure the oscillation frequency 
$\Delta m_d$.
By now the $\Delta m_d$ measurements,
as well as the top quark mass measurements,
have become sufficiently precise
that 
other uncertainties, in particular the $B^0$ meson decay constant, 
limit the 
precision of the extraction of $|V_{td}|$.

The same phenomenon 
of particle-antiparticle oscillations 
is expected for 
the $B^0_s \overline{B}$$^0_s$ system, where
the 
relevant 
element of the KM 
matrix is $|V_{ts}|$.
Due to the difference in the involved matrix elements,
$B^0_s \overline{B}$$^0_s$ mixing is expected to
proceed with a higher oscillation frequency, 
and so far only lower limits on the frequency
$\Delta m_s$ 
have been 
placed~\cite{new_LEP_common,new_Bs_limits}.
Once 
the $B^0_s \overline{B}$$^0_s$ oscillation is established, 
a measurement of the ratio
of the two oscillation frequencies,
$\Delta m_s / \Delta m_d$,
would provide a useful constraint on the
ratio of the 
KM matrix elements $|V_{ts}| / |V_{td} | $
with less theoretical uncertainty.

In this paper 
we report a measurement of 
$B^0 {\overline B}$$^0$  mixing
using partially reconstructed semileptonic decays.
The data used in this analysis
were collected 
in 1992--1995
with the 
CDF detector 
at the Fermilab Tevatron proton-antiproton
collider 
at 
a center-of-mass energy of $\sqrt{s}=1.8$ TeV,
and correspond to an integrated luminosity of about 110 pb$^{-1}$.
We use a data sample where events are 
collected 
on
the 
existence of two lepton candidates.
In order to identify semileptonic decays of ${\overline B}$ mesons,
we select events with a lepton ($e^-$ or $\mu^-$, denoted by $\ell^-$)
associated with a $D^{*+}$ meson.
(Throughout this paper 
a reference to a particular charge state
also implies its charge conjugate.)
The $\ell^- D^{*+}$ pairs 
consist mostly of ${\overline B}$$^0$ decays.
The $D^{*+}$ decays are reconstructed 
using the decay mode $D^{*+} \rightarrow D^0 \pi^+$, followed by
$D^0 \rightarrow K^- \pi^+$, 
$K^- \pi^+ \pi^+ \pi^-$, 
or $K^- \pi^+ \pi^0$.
About 500 such decays are reconstructed in the data sample.
We reconstruct their decay vertices 
and estimate
the proper decay length of the 
${\overline B}$$^0$ meson using   
the 
momentum
of the $\ell^- D^{*+}$ system.
The charge of the final state 
lepton 
identifies the flavor of
the ${\overline B}$$^0$ meson at the time of its decay
($\ell^- D^{*+}$ for ${\overline B}$$^0$, and  $\ell^+ D^{*-}$ for $B^0$).
The ${\overline B}$$^0$ meson flavor at its production
is inferred from the charge of the second lepton
in the event (${\bar b} \rightarrow B \rightarrow \ell^+ \nu X$),
assuming that $b$ and $\bar b$ quarks are produced in pairs.
Thus, 
in ideal cases, 
an opposite-sign lepton pair identifies
an unmixed decay, 
and a same-sign pair identifies a mixed decay.
We examine decay length distributions
of opposite-sign and same-sign events
and extract the oscillation frequency $\Delta m_d$.

%
\section{CDF Detector and Trigger}
The CDF detector is described in detail elsewhere~\cite{CDF}. 
We describe here only the detector components 
most relevant to this analysis. 
Inside the 1.4 T solenoid
the silicon vertex detector (SVX)~\cite{svx} 
and the central tracking chamber (CTC)
provide the tracking and momentum analysis of charged particles.
The CTC is a cylindrical drift chamber containing 84 measurement layers.
It 
covers the pseudorapidity interval $|\eta| < 1.1$, 
where $\eta=-\ln[\tan(\theta/2)]$. 
In CDF, $\varphi$ is the azimuthal angle, 
$\theta$ is the polar angle measured from the proton direction, 
and $r$ is the radius from the beam axis ($z$-axis). 
The SVX consists of four layers of silicon micro-strip detectors
located at radii between  2.9 
and 7.9~cm from the beam line 
and provides spatial measurements in the $r$-$\varphi$ plane
with a resolution of 13 $\mu$m. 
It
gives
a track impact parameter resolution of about 
$(13 + 40/p_T)~\mu$m~\cite{svx}, 
where $p_T$ is the transverse momentum of the track 
with respect to the beam axis  and    
measured in GeV/$c$.
The silicon detectors extend to $\pm~25$ cm along the $z$ axis, 
where $z$ is parallel to the proton beam axis. 
Since the vertex distribution for ${\bar p} p$ collisions
has an rms width of $\pm 30$~cm along the $z$ direction, 
a substantial fraction of the interactions occurs outside
of the SVX coverage; 
as a result, the average geometric acceptance of the SVX 
is about 60\%.
The transverse profile of the Tevatron beam
is circular and has an rms spread 
along both $x$ and $y$ axes
of 
$\sim 35$ $\mu$m for the data taking period in 1992--1993
and $\sim 25$ $\mu$m in 1994--1995.
The $p_T$ resolution of the CTC combined with the SVX is 
$\sigma(p_T)/p_T = [ (0.0066)^2 + (0.0009 \, p_T)^2 ]^{1/2} $.  
Electromagnetic (CEM) 
and hadronic (CHA) 
calorimeters
with projective tower geometry 
are located 
outside the solenoid
and
cover the pseudorapidity region $|\eta|<1.1$,
with a segmentation of $\Delta \varphi = 15^{\circ}$
and $\Delta \eta \simeq 0.11$.
A layer of proportional chambers (CES)  is embedded near shower
maximum in the CEM and 
provides a more precise measurement of electromagnetic 
shower profiles and 
an
additional measurement of pulse height.
A layer of proportional chambers (CPR) is 
also 
installed between the solenoid and the CEM 
and samples the electromagnetic showers at about one radiation length.
Two muon subsystems 
in the central rapidity region 
$( | \eta | < 0.6 )$  
are used for muon identification:
the central muon chambers (CMU) 
located just behind the CHA calorimeter,
and the central upgrade muon chambers (CMP) 
which 
lie behind an 
additional 60~cm of steel.
The central muon extension chambers (CMX),
covering a 
rapidity region 
up to $( | \eta | < 1.0  )$,  
are 
also 
used.

CDF uses a three-level trigger system, where 
at the first two levels decisions are made with dedicated hardware.
The information available at this stage
includes energy 
deposits 
in the CEM and CHA calorimeters,
high $p_T$ tracks found in CTC by a track processor,
and track segments found in the muon subsystems.
At the third level of the trigger, 
the 
event selection is based on a
version of off-line reconstruction programs
optimized for speed.
The lepton selection criteria used in level 3 
are similar to those described  in the next Section.

Events containing semileptonic $B$ decays 
and used for this analysis  
are collected using 
two triggers that require two lepton candidates in
an event.
The first trigger requires 
both 
an electron candidate and a muon candidate.
The $E_T$ 
threshold for the 
electron 
is 5~GeV,
where $E_T \equiv E\sin \theta$,
and $E$ is the energy measured in the CEM.
In addition, 
a track is required
in the CTC with $p_T > 4.7$ GeV/$c$
that points at the calorimeter tower in $\varphi$.
The muon candidate requires 
a track in the CTC
with matched track segments in the CMU or CMX system
corresponding to 
a particle with  $p_T > 2.7$ GeV/$c$.
The second trigger requires two muon candidates,
where the $p_T$ threshold is 2.2 GeV/$c$ for each muon,
and at least one of the muons is required to have track segments
in both the CMU and CMP chambers.

\section{Reconstruction of Semileptonic Decays of $B$ Mesons}
The analysis starts with identification of lepton candidates.
We require at least two good lepton candidates 
in an event.
We then look for the charm meson $D^{*+}$ near each lepton
candidate to identify the ${\overline B}$ meson decay 
${\overline B} \rightarrow \ell^- \bar {\nu} D^{*+} X$.
A proper correlation between the lepton charge
and the charm flavor, namely
$\ell^-$ with $D^{*+}$, 
and 
not $\ell^+$ with $D^{*+}$,
is required.
This decay is used to measure the proper decay length
of the ${\overline B}$$^0$ meson and to identify the decay flavor.
The charge of  
the other lepton candidate in the event is used to 
infer 
the flavor of the ${\overline B}$$^0$ meson at its production.
\subsection {Lepton identification}
The identification of electrons makes use of information 
from both calorimeters 
and tracking chambers.
We require the following:\\ 
\begin{itemize}
\item Longitudinal 
shower
profile consistent with 
electrons, 
{\em i.e.}, 
small leakage 
of 
energy 
into 
the CHA.
\item Lateral shower profiles measured with the CEM~\cite{lateral1}
and the CES~\cite{lateral2} consistent with test beam data. 
\item Association of a high $p_T$ track with the calorimeter
shower based on position matching and energy-to-momentum ratio. 
\item Pulse heights in the CES and CPR consistent with an electron.
\end{itemize}
Photon conversion electrons, 
as well as the Dalitz decays of $\pi^0$ mesons,
are removed by looking for oppositely charged tracks that 
have
small opening angles with the 
electron candidate.

Muons are identified based on the geometrical 
match 
between the track segments in the muon chambers
and an extrapolated CTC track.
We compute the $\chi^2$ of the matching, where the uncertainty is
dominated by multiple Coulomb scattering in the detector material.
We require $\chi^2 < 9$ in the $r$-$\varphi$ view (CMU and CMP)
and $\chi^2 < 12$ in the $r$-$z$ view (CMU).
For muon candidates in the CMX we require $\chi^2 < 9$ 
in both
the $r$-$\varphi$ and $r$-$z$ views.


\subsection {Charm meson reconstruction}



To identify  ${\overline B} \rightarrow \ell^- {\bar \nu}D^{*+} X$ candidates, 
we search for $D^{*+} \rightarrow D^0 \pi^+$ decays
in the vicinity of a lepton candidate
using 
two fully reconstructed $D^0$ decay modes,
$D^0 \rightarrow K^- \pi^+$ and
$D^0 \rightarrow K^- \pi^+ \pi^+ \pi^-$, and
one partially reconstructed mode,
$D^0 \rightarrow K^- \pi^+ \pi^0$.
To reconstruct $D^0 \rightarrow K^- \pi^+$ decays,
we first select oppositely charged pairs of particles using CTC tracks,
where
the kaon mass is assigned to the particle with the
same charge as the lepton,
as is the case in semileptonic $B$ decays.
The kaon (pion) candidate is then required to have
transverse momentum above 1.2 (0.4) GeV/$c$,
and to be within a cone of radius $\Delta R = 0.8$ (1.0) around the lepton
in $\eta$-$\varphi$ space,
where 
$\Delta R = [ (\Delta \eta)^2 + (\Delta \varphi)^2 ] ^{1/2} $.
To ensure accurate decay length measurement, 
each candidate track,
as well as the lepton track,
is required to be reconstructed in 
the 
SVX
with hits in at least two layers out of 
the 
possible four, 
and with $\chi^2 < 6$ per hit.
To reduce combinatorial background,
we require the decay vertex of the $D^0$ candidate
to be positively displaced 
along its flight direction  in  the transverse plane
with respect to 
the position of the  primary vertex.
The primary vertex is approximated 
by the position of the Tevatron beam,
which has been determined 
using independent events~\cite{cdf_psi_life}.
For the $D^0 \rightarrow K^- \pi^+ \pi^+ \pi^-$ mode,
the kaon (pion) candidate is required to have
transverse momentum above 1.2 (0.5) GeV/$c$,
and to be within a cone of radius 
$\Delta R = 0.6$ 
(1.0) around the lepton candidate.
For the  $D^0 \rightarrow K^- \pi^+ \pi^0$ mode,
the kaon (pion) candidate is required to have
transverse momentum above 1.2 (0.4) 
GeV/$c$,
and to be within a cone of radius 
$\Delta R = 0.7$  
(0.8) 
around the lepton candidate.

\begin{figure}[p]
\vspace{1in}
\epsfysize=5.5in
\epsffile[ 30 144 522 648]{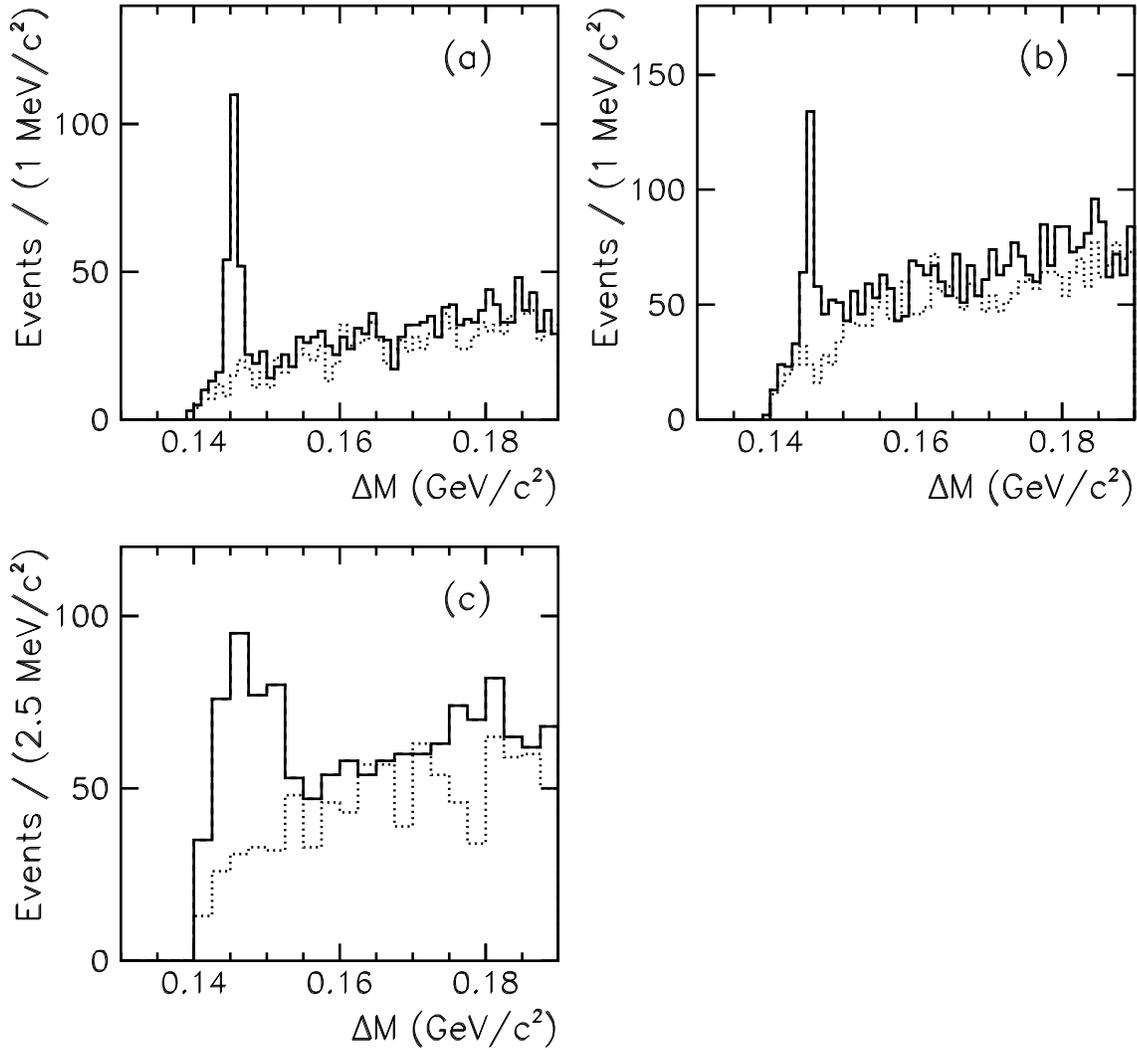}	
\vspace{0.5in}
\caption{ Reconstructed $D^{*+} \rightarrow D^0 \pi^+$ signals 
in events with two lepton candidates. The $D^{*+}$ meson
is associated with a lepton ($\ell^-$) candidate.
Distributions of $\Delta M$ for three $D^0$ decay modes are shown: 
(a) $D^0 \rightarrow K^- \pi^+$,
(b) $D^0 \rightarrow K^- \pi^+ \pi^+ \pi^-$,  and
(c) $D^0 \rightarrow K^- \pi^+ \pi^0$.
Dotted 
histograms show 
the distributions for
wrong sign ($D^0 \pi^-$) combinations.
}
\label {fig:signal}
\end {figure}

In order to qualify as a candidate for the signal,
the $D^0$ candidate has to be in the mass range 
1.83 to 1.90 GeV/$c^2$ 
for the fully reconstructed mode $D^0 \rightarrow K^- \pi^+$
and 
in the range 
1.84 to 1.88 GeV/$c^2$
for the 
$\rightarrow K^- \pi^+ \pi^+ \pi^-$ mode.
For the partially reconstructed mode $D^0 \rightarrow K^- \pi^+ \pi^0$, 
we require the mass of a $K^- \pi^+$ pair
to be between 1.5 and 1.7 GeV/$c^2$;
we do not reconstruct the $\pi^0$ meson and
in the subsequent analysis
treat the $K^-\pi^+$ pair as if it were a $D^0$ meson.
For each mode, we reconstruct the $D^{*+}$ meson
by combining an additional CTC track,
assumed to have the pion mass, with the $D^0$ candidate,
and computing the mass difference, $\Delta M$, between the
$D^0 \pi^+$ and $D^0$ candidates.
Figure~\ref{fig:signal} 
shows 
 the $\Delta M$ distributions
for the three $D^0$ decay modes. 
In Fig.~\ref{fig:signal}(c)
the peak is broadened because of the missing $\pi^0$ meson.
The dotted 
histograms 
show 
the spectra from the ``wrong sign" ($D^0 \pi^-$) combinations,
where no significant signals are observed.
We define the signal region as follows: 
the two fully reconstructed modes
use the $\Delta M$ range 0.144 to 0.147 GeV/$c^2$, and
the $K^- \pi^+ \pi^0$ mode uses the 
range $\Delta M <0.155$ GeV/$c^2$. 
The numbers of events in the signal regions are
216, 256, and 416 for the three modes.
We estimate the numbers of combinatorial background events 
by using the shapes of the $\Delta M$ spectra of 
the wrong sign ($D^0 \pi^-$) combinations
and normalizing them 
to the number of events 
in the $\Delta M$ sideband.
The estimated background fractions    
are
$0.227 \pm 0.036$, $0.326 \pm 0.040$ and $0.543 \pm 0.050$, respectively.
They are summarized in Table~\ref{tb:signal}.


\subsection {Sample composition}

Apart from combinatorial backgrounds,
the $\ell^- D^{*+}$ signal sample contains
events which originated from
physics sources other than the ${\overline B}$$^0$ meson decays.
The main contribution comes from 
$B^-$ meson decays.
The semileptonic decays of 
$B$ mesons
can be expressed as
${\overline B} \rightarrow \ell^- {\overline \nu} {\bf D}$, where
${\bf D}$ is a charm system whose charge 
is correlated with the $B$ meson charge.
If only the two lowest mass 
charm states, 
pseudoscalar ($D$) and vector ($D^*$) mesons,
are produced,
the $\ell^- D^{*+}$ combination
can arise only from the ${\overline B}$$^0$ decay.
However, it has been known that the above two lowest mass 
states
do not saturate the total semileptonic decay rates.
All data indicate
that higher mass charm mesons, $D^{**}$ states,
as well as non-resonant $D^{(*)} \pi$ pairs,
are responsible for the rest of the
semileptonic decays~\cite{PDG98}. 
In this analysis 
we do not distinguish resonant and non-resonant components,
and refer to both of them as $D^{**}$ 
mesons.  

These $D^{**}$ meson decays can dilute the charge correlation
between the final states and the parent $B$ meson.
For example, 
the $D^{**0}$  meson 
can be produced by the decay
$B^- \rightarrow \ell^- {\bar \nu} D^{**0}$,
which subsequently can
produce both $D^{*+} \pi^-$ and $D^{*0} \pi^0$ final states.
This results
in misidentification of the $B^-$ meson 
decay 
as
${\overline B}$$^0 \rightarrow D^{*+} \ell^- {\bar \nu} X$.
Nevertheless, 
the $\ell^- D^{*+}$ combination is dominated by 
${\overline B}$$^0$ meson decays. 

In order to estimate the fraction $g^-$ of 
$B^-$ decays
relative to the sum of $B^-$ and ${\overline B}$$^0$ mesons
in the observed $\ell^- D^{*+}$  sample, 
we follow the method used in 
the 
CDF 
measurement of
the $B^-$ and ${\overline B}$$^0$ meson 
lifetimes~\cite{my_lifetimes} using semileptonic decays.
We describe the method here as well.

The production rates of charged and neutral $B$ mesons
and their semileptonic decay widths are assumed to be equal. 
We also assume the $D^{**}$ mesons decay exclusively to a $D^{(*)}\pi$ pair
via the strong interaction, thereby allowing us to 
determine the branching fractions, 
{\em e.g.}~$D^{(*)+} \pi^0$ vs.~$D^{(*)0} \pi^+$, 
using isospin symmetry. 
We consider 
three 
factors affecting the composition.
First, the composition depends on
the fraction $f^{**}$
of the $D^{**}$ mesons produced
in semileptonic $B$ decays,
\[
f^{**} \equiv
\frac { {\cal B} ({\overline B} \rightarrow \ell^- {\bar \nu} D^{**}) }
 { {\cal B} ({\overline B} \rightarrow \ell^- {\bar \nu} D X ) }
= 1 - \frac
 { {\cal B} ({\overline B} \rightarrow \ell^- {\bar \nu} D ) 
 + {\cal B} ({\overline B} \rightarrow \ell^- {\bar \nu} D^{*} ) }
 { {\cal B} ({\overline B} \rightarrow \ell^- {\bar \nu} D X )   },
\]
where ${\cal B}$ denotes a branching fraction
and ${\overline B}$ is a $B^-$ or ${\overline B}$$^0$ meson.
The CLEO experiment measures the fraction of exclusive
decays to the two lowest mass states to be 
$0.64 \pm 0.10 \pm 0.06$~\cite{CLEO}.
Thus, 
we estimate 
that $f^{**} = 0.36 \pm 0.12$.
A few experiments have recently observed
some $D^{**}$ modes~\cite{ddst}, 
but the sum of exclusive modes
still 
does not equal the total semileptonic rate.
Second, 
the fraction   
$g^-$ depends on the relative abundance of 
various 
possible $D^{**}$ states, because
some of them decay only to $D^*\pi$ and others to $D\pi$,
depending on the spin and parity.
The abundance is not measured very well at present.
Changing the abundance is equivalent to changing the
branching fractions 
for $D^* \pi$ and $D \pi$
modes 
averaged over various $D^{**}$ states. 
We define 
the 
quantity    
\[
P_V \equiv 
\frac { {\cal B} (D^{**}\to D^*\pi) } 
      { {\cal B} (D^{**}\to D^* \pi) + {\cal B} (D^{**}\to D \pi)  } .
\]
We assume the relative abundance 
of the four $D^{**}$ mesons
predicted
by 
the Isgur-Scora-Grinstein-Wise (ISGW) model~\cite{ISGW},
which corresponds to 
$P_V = 0.64$.   
After inclusion of non-resonant contributions,
we use $P_V = 0.65$ as our nominal choice.  
We also consider the values $P_V = 0.26$ and $1.00$.
Third, 
the composition depends on the ratio
of the $B^-$ and ${\overline B}$$^0$ meson lifetimes,
because the number of $\ell^- D^{*+}$ events 
is proportional to the semileptonic branching fraction, 
which is the product of the lifetime and the partial width.
We use the ratio
$\tau(B^-) / \tau ( {\overline B}$$^0) = 1.02 \pm 0.05$~\cite{PDG}.

We also 
take into account the differences in the reconstruction
efficiencies for
the  ${\overline B} \rightarrow \ell^- {\bar \nu} D^*$ and $D^{**}$
decay modes. We examine this effect by using 
Monte Carlo events
where the ISGW model is used to describe the semileptonic decays.
We shall describe the Monte Carlo 
simulation 
later.
We find that 
the 
$D^{**}$
mode has an efficiency that is lower 
than in the $D^*$ mode
by about 50\% (25\%)
for leptons above 5 GeV/$c$ (2 GeV/$c$).  

We find that
$g^- = 0.19 \, ^{+0.08} _{-0.10}$ for the $\mu^- D^{*+}$ sample
and $g^- = 0.14 \, ^{+0.06} _{-0.08}$ for the $e^- D^{*+}$ sample.
The central values correspond to the nominal choice
of the parameters, $f^{**} = 0.36$, $P_V = 0.65$, 
and 
$\tau(B^-) / \tau ( {\overline B}$$^0) = 1.02$.
The 
uncertainties reflect maximum changes in $g^-$
when $f^{**}$, $P_V$ and the lifetime ratio
are changed within 
their uncertainties, 
namely $f^{**}$ to 0.24 and 0.48,
$P_V$ to 0.26 and 1.0, and 
$\tau(B^-) / \tau ( {\overline B}$$^0)$ to 0.97 and 1.07.
The difference between the muon and electron channels
arises from the difference in kinematic requirements.

There are other physics processes that can produce the lepton-$D^{*+}$
signature. 
The largest background comes from the decay
of the ${\overline B}$$^0_s$ meson, 
${\overline B}$$^0_s  \rightarrow \ell^- {\overline \nu} D_s^{**+}$,
followed by $D_s^{**+} \rightarrow D^{*+} K^0$.
This process is estimated to contribute 
about 3\% of the
lepton-$D^{*+}$ signal.
Other processes such as 
${\overline B} \rightarrow \tau^- {\bar \nu}_\tau D^{*+} X$
followed by $\tau^- \rightarrow \ell^- {\bar \nu}_\ell \nu_{\tau}$,
and 
${\overline B} \rightarrow D_s^{-} D^{*+} X$
followed by $D_s^{-} \rightarrow \ell^- X$,
are suppressed severely   because of 
branching fractions and kinematic requirements on leptons.
We ignore    
these backgrounds here.
Therefore, the fraction of 
${\overline B}$$^0$ mesons is given by $g^0 = 1 - g^-$.
We treat effects of the physics backgrounds
as a systematic uncertainty.

\section {Decay length measurement and momentum estimate}
\begin{figure}[p]
\vspace{1.5in}
\epsfysize=6.0in
\epsffile[ 50 144 522 648]{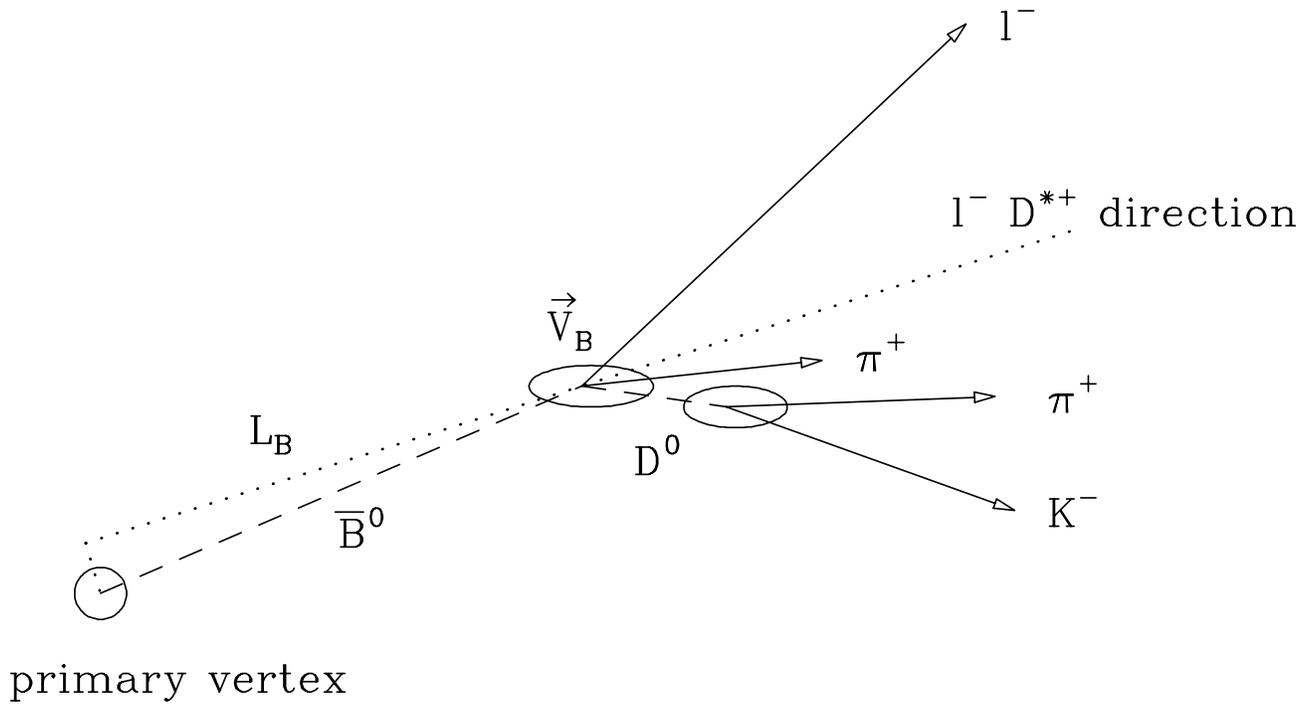}
\vspace{1in}
\caption
{Schematic representation of the decay
${\overline B}$$^0 \rightarrow 
\ell^- {\bar \nu} D^{*+}$, 
followed by 
$D^{*+} \rightarrow D^0 \pi^+$ and 
$D^0 \rightarrow K^- \pi^+$.
}
\label {fig:schematics}
\end {figure}

The $B$ meson decay vertex ${\vec V}_B$
is obtained by
intersecting the trajectory of the lepton track with the flight path
of the $D^0$ candidate. 
The $B$ decay length $L_B$ is defined as the displacement 
of ${\vec V}_B$ from the primary vertex ${\vec V}_P$, 
measured in the plane perpendicular to 
the beam axis,
and
projected onto the transverse momentum vector
of the lepton-$D^{*+}$ system:
\[ L_B \equiv
\frac { ( {\vec V}_B - {\vec V}_P ) \cdot {\vec p}_T^{\, \ell^- D^{*+} } }
      { p_T^{ \, \ell^-D^{*+} } }.
\]
A schematic representation of the 
${\overline B}$$^0$ meson semileptonic decay
is illustrated in Fig.~\ref{fig:schematics}.

To measure 
the 
proper decay length 
of a $B$ meson, 
we need to know the momentum of the $B$ meson.
In semileptonic decays, 
the $B$ meson momentum cannot be measured precisely
because of the missing neutrino.
We use 
the transverse momentum of the observed system, 
$p_T^{\ell^-D^{*+}}$, 
to estimate the $B$ meson transverse momentum $p_T^B$ for each event. 
We denote the ratio of the two momenta by 
$K \equiv  p_T^{\, \ell^- D^{*+}}  / p_T^B$,
and introduce a corrected decay length defined as
\[
  x \equiv L_B \frac {m_B}  {p_T^{\, \ell^- D^{*+}} } \langle K \rangle ,
\]
which we call the ``pseudo-proper decay length.''
The average correction for the missing momentum is achieved by
the constant $\langle K \rangle$.
The correction for a finite width of
the distribution of the ratio $K$	
is performed 
during fits to decay length distributions.
We shall describe the fits later.

A typical resolution on this decay length $x$ due to vertex determination
is 50 $\mu$m,
including the contribution from the finite size of the primary vertex.
For subsequent decay length measurements, we use only those events
in which 
the resolutions on reconstructed decay lengths 
are 
smaller than 0.05 cm. 
We also 
require the proper decay length of 
the $D^0$ meson, measured from the $B$ meson decay vertex to
the $D^0$ decay vertex, 
to be in the range from $-0.1$~cm to 0.1~cm, 
with its uncertainty smaller than 0.05~cm.
These cuts reject 
poorly measured decays and 
reduce 
random track combinations.
In addition, we limit ourselves to events with
reconstructed decay lengths 
in the range between $-0.15$ cm and 0.3 cm.
These cuts have been applied already 
for the charm signals shown in Fig.~\ref{fig:signal}.

\begin{figure}[p]
\vspace{2.0in}
\epsfysize=6.5in
\epsffile[ 50 144 522 648]{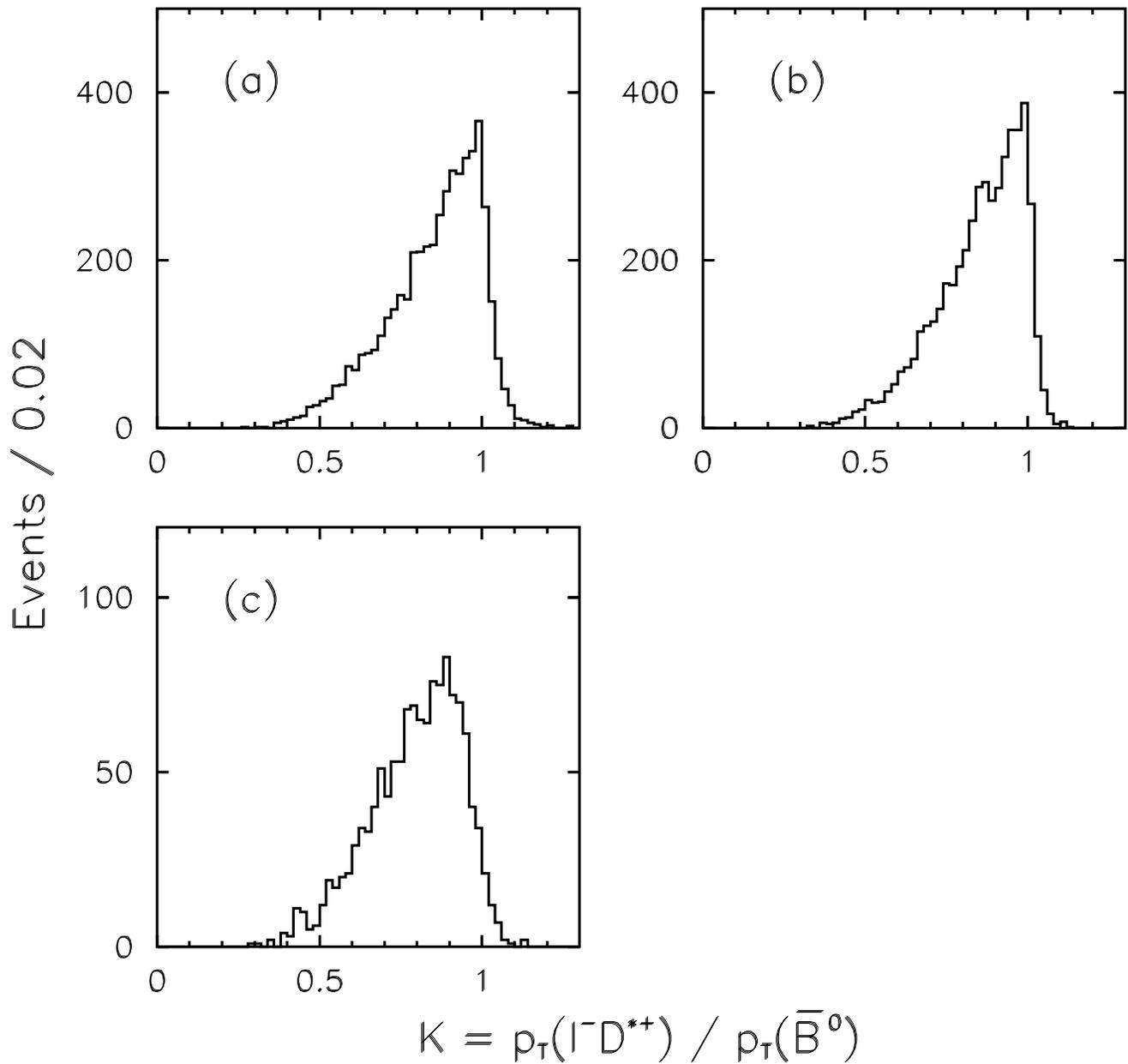}  
\vspace{1in}
\caption
{Distributions of the
momentum ratio $K$ (see text)
obtained from Monte Carlo calculations
for decays
${\overline B}$$^0 \rightarrow 
\ell^- {\bar \nu} D^{*+} X$,  
followed by 
$D^{*+} \rightarrow D^0 \pi^+$.
Three $D^0$ decay modes are shown:
(a) $D^0 \rightarrow K^- \pi^+$,
(b) $D^0 \rightarrow K^- \pi^+ \pi^+ \pi^-$,  and
(c) $D^0 \rightarrow K^- \pi^+ \pi^0$.
}
\label {fig:k}
\end {figure}

The distribution of the momentum ratio
$K$		
is obtained 
from a Monte Carlo calculation.
The $b$ quarks are generated 
according to the $p_T$ spectrum 
by the QCD calculation in the next-to-leading order~\cite{NDE}.
The fragmentation model by Peterson and others~\cite{Peterson} is used.
The CLEO event generator~\cite{QQ} is 
then 
used to describe 
the 
$B$ meson decays. 
In particular, the semileptonic
decays adopt the ISGW model~\cite{ISGW}.
A typical $K$ distribution thus obtained
has an average value of 0.85 with an rms width of 0.14,
and shows only a weak dependence on
$p_T^{\ell^- D^{*+}}$
in the range of interest,
which is typically between 10 and 20 GeV/$c^2$.
It is also independent of 
the $D^0$ decay mode except for the
partially reconstructed mode $D^0 \rightarrow K^- \pi^+ \pi^0$,
which has a slightly 
lower mean value 
(about 0.80) 
because of the missing $\pi^0$ meson.
Typical $K$ distributions are shown in
Fig.~\ref{fig:k}.

We fit the observed pseudo-proper decay length distributions 
for both opposite-sign and same-sign events.
This fit determines parameters that will be used later
in the fit for the oscillation frequency $\Delta m_d$.
It also yields the $B^0$ meson lifetime 
as a check of the momentum correction described above.
We use the maximum likelihood method.
The likelihood  
used to fit the events in the signal region 
is expressed as 
\begin{displaymath}
{\cal L}_{\rm SIG}  =  \prod_i [ (1-f_{\rm BG}) {\cal F}_{\rm SIG}(x_i)
 + f_{\rm BG} {\cal F}_{\rm BG} (x_i) ],
\end{displaymath}
where 
$x_i$ is the pseudo-proper decay length measured for 
event $i$, 
and the product is taken over observed events in the sample.
The first term in the likelihood function
represents 
the contribution of 
$B$ decay signal events, while
the second term accounts for combinatorial background events
whose fraction in the sample is $f_{\rm BG}$.

The signal probability density function  ${\cal F}_{\rm SIG}(x)$
has two components   
and is expressed as 
\[
{\cal F}_{\rm SIG}(x) = g^-  {\cal F}_{\rm SIG}^- (x)
                 + (1-g^-) {\cal F}_{\rm SIG}^0 (x),
\]
where 
${\cal F}_{\rm SIG}^- (x)$ and ${\cal F}_{\rm SIG}^0 (x)$ 
are the normalized probability density functions for
the $B^-$ and ${\overline B}$$^0$ meson decays, respectively,
and $g^-$ is the fraction of $B^-$ mesons as defined earlier.
Each component 
consists of 
an exponential decay function, 
defined for positive decay lengths,
smeared with a normalized $K$ distribution $D(K)$
and a Gaussian distribution with width $s \sigma_i$:
\begin{equation}
	{\cal F}_{\rm SIG}^{-,0} (x) = \int dK \, D^{-,0}(K)
\, 
\left[
\theta(x) \,
\frac{K} { c \tau  \langle K \rangle } 
\exp \left( - \frac{Kx} { c \tau \langle K \rangle } \right)
\otimes G(x)
\right]
,
\label{eq:life_signal}
\end{equation}
where 
$\tau$ is the 
appropriate 
$B$ meson lifetime, $c$ is the speed of light,
$\theta(x)$ is the step function defined as 
$\theta(x)=1$ for $x \geq 0$ and 
$\theta(x)=0$ for $x  < 0$,
and the symbol ``$\otimes$" denotes a convolution.
$G(x)$ is the Gaussian distribution given by
\[
	G(x) = \frac{ 1} { s \sigma_i \sqrt{ 2 \pi } }
	  \, \exp \left( - \frac { x^2 } { 2 s^2 \sigma_i^2 } \right),
\]
where 
$\sigma_i$ is the estimated resolution on $x_i$.
The scale factor $s$ is introduced 
in order to account for a possible incompleteness
of our estimate of the decay length resolution.
The integration over the momentum ratio $K$ 
is
approximated by 
a finite sum
\[
\int dK D(K) \rightarrow \sum_j D(K_j) \Delta K,
\]
where the sum is taken over bin $j$
of a histogrammed distribution $D(K_j)$
with bin width $\Delta K = 0.02$.
The $K$ distributions  for $B^-$ and ${\overline B}$$^0$ mesons 
are slightly different 
because 
the $B^- \rightarrow \ell^- {\bar \nu}  D^{*+} X$ decay
involves 
more missing particles.

The pseudo-proper decay length 
distribution
of combinatorial background events, ${\cal F}_{\rm BG}(x)$, 
is measured 
using 
$\Delta M$ 
sideband events, 
assuming that
they represent the combinatorial background events under 
the 
signal mass peaks.
The functional form of the distribution is
parameterized empirically
by a sum of a Gaussian distribution centered at zero,
and positive and negative exponential tails 
smeared with a Gaussian distribution:
\begin {eqnarray}
{\cal F} _{\rm BG} ( x )  & = & 
 (1-f_- - f_+) \,G( x )   \nonumber \\
& + & 
\frac { f_+ } { \lambda_+ }
\, \theta(x) \exp 
\left ( - \frac {x }  {\lambda_+ } \right ) 
\otimes G ( x ) 
\nonumber \\
&  + & 
\frac { f_- } { \lambda_- }
\, \theta(-x) \exp 
\left ( + \frac {x }  {\lambda_- } \right ) 
\otimes G ( x ).
\label{eq:life}
\end {eqnarray}


The shape of the background function 
(parameters $f_{\pm}$ and $\lambda_{\pm}$)
and the 
resolution 
scale factor $s$,
as well as the 
$B$ meson
lifetime $c \tau$,  
are determined from a simultaneous fit 
to 
signal and sideband events.
We fix the ratio of the $B^-$ and ${\overline B}$$^0$ meson 
lifetimes 
and fit for the ${\overline B}$$^0$ meson lifetime only.
To determine those parameters,
we use the combined likelihood ${\cal L}$ defined as
${\cal L} =  {\cal L}_{\rm SIG} \; {\cal L}_{\rm BG} $,
where ${\cal L}_{\rm BG}  =  \prod_k {\cal F}_{\rm BG} (x_k)$
and  the product 
is taken over event $k$ in the background sample.
The amount of combinatorial background $f_{\rm BG}$ is also
a  parameter in the simultaneous fit.
This parameter is constrained by adding a term
$\frac{1}{2} \chi^2= \frac{1}{2} 
(f_{\rm BG}- \langle f_{\rm BG} \rangle)^2 / \sigma_{\rm BG}^2$  to
the negative log-likelihood $- \ell = - \ln {\cal L}$.
The average background fraction $\langle f_{\rm BG} \rangle$
and its uncertainty $\sigma_{\rm BG}$
are
estimated from 
the signal mass distributions 
and 
are given in Table~\ref{tb:signal}.

The background sample 
for the $\ell^- D^{*+}$ candidates
is taken from the $\Delta M$ sidebands:
we use the right sign ($D^0 \pi^+$) sideband
$0.15 < \Delta M < 0.19$ GeV/$c^2$
for the two fully reconstructed $D^0$ modes,
and $0.16 < \Delta M < 0.19$ GeV/$c^2$ 
for the $D^0 \rightarrow K^- \pi^+ \pi^0$ mode.
We also use the wrong sign pion combinations 
in the range $\Delta M < 0.19$ GeV/$c^2$ for all three $D^0$ decay modes.
The background samples are summarized in Table~\ref{tb:control}.

\begin{figure}[p]
\vspace{2.0in}
\epsfysize=6.5in
\epsffile[ 50 144 522 648]{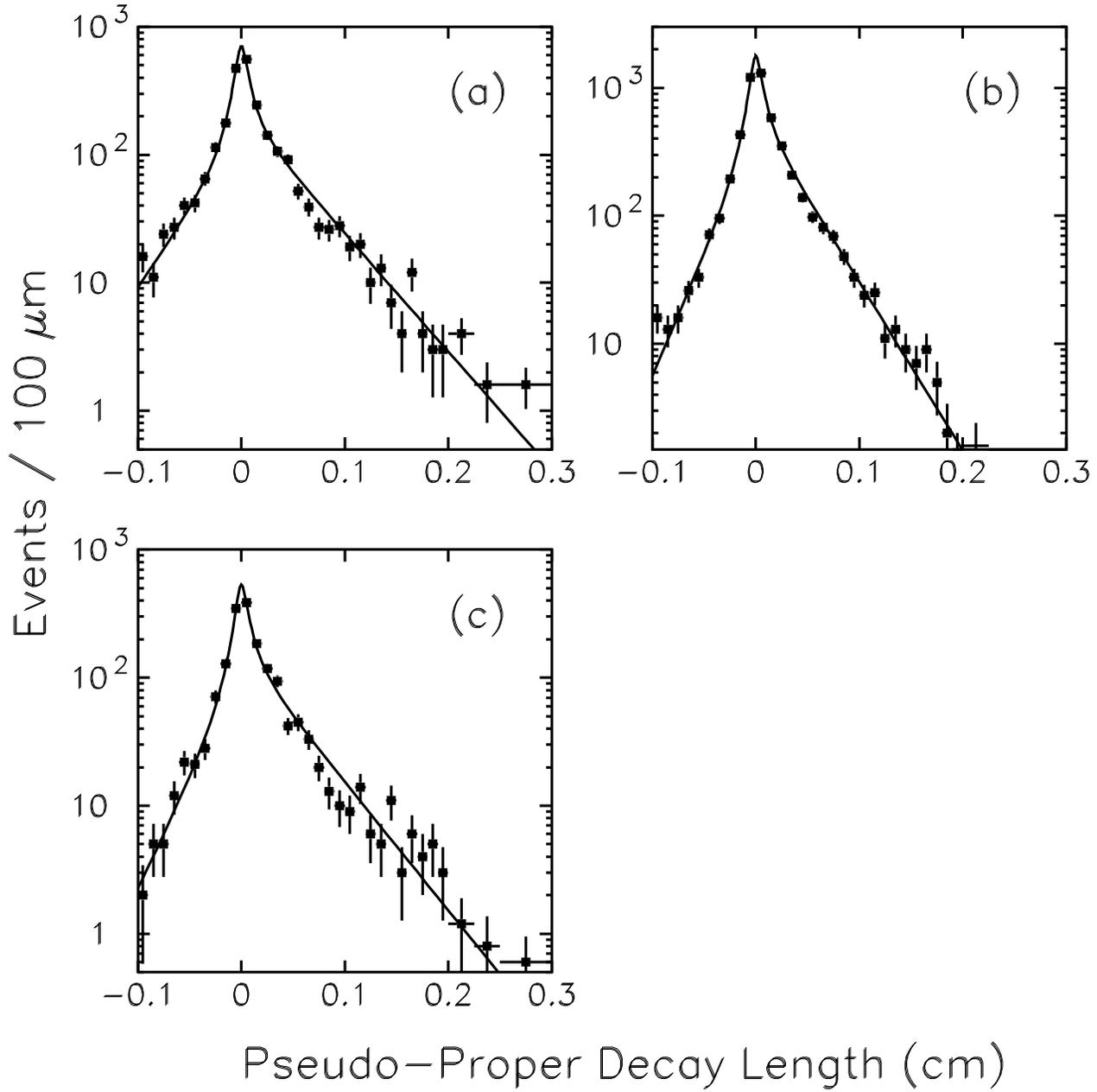}  
\vspace{1in}
\caption
{Distributions of $B$ meson pseudo-proper decay lengths
for $\ell^- D^{*+}$ background samples (points).
Three decay modes are shown:
(a) $D^{*+} \rightarrow D^0 \pi^+$, $D^0 \rightarrow K^- \pi^+$,
(b) $D^{*+} \rightarrow D^0 \pi^+$, $D^0 \rightarrow K^- \pi^+ \pi^+ \pi^-$, 
and
(c) $D^{*+} \rightarrow D^0 \pi^+$, $D^0 \rightarrow K^- \pi^+ \pi^0$.
Curves show fit results.
}
\label {fig:bg}
\end {figure}

\begin{figure}[p]
\vspace{2.0in}
\epsfysize=6.5in
\epsffile[ 50 144 522 648]{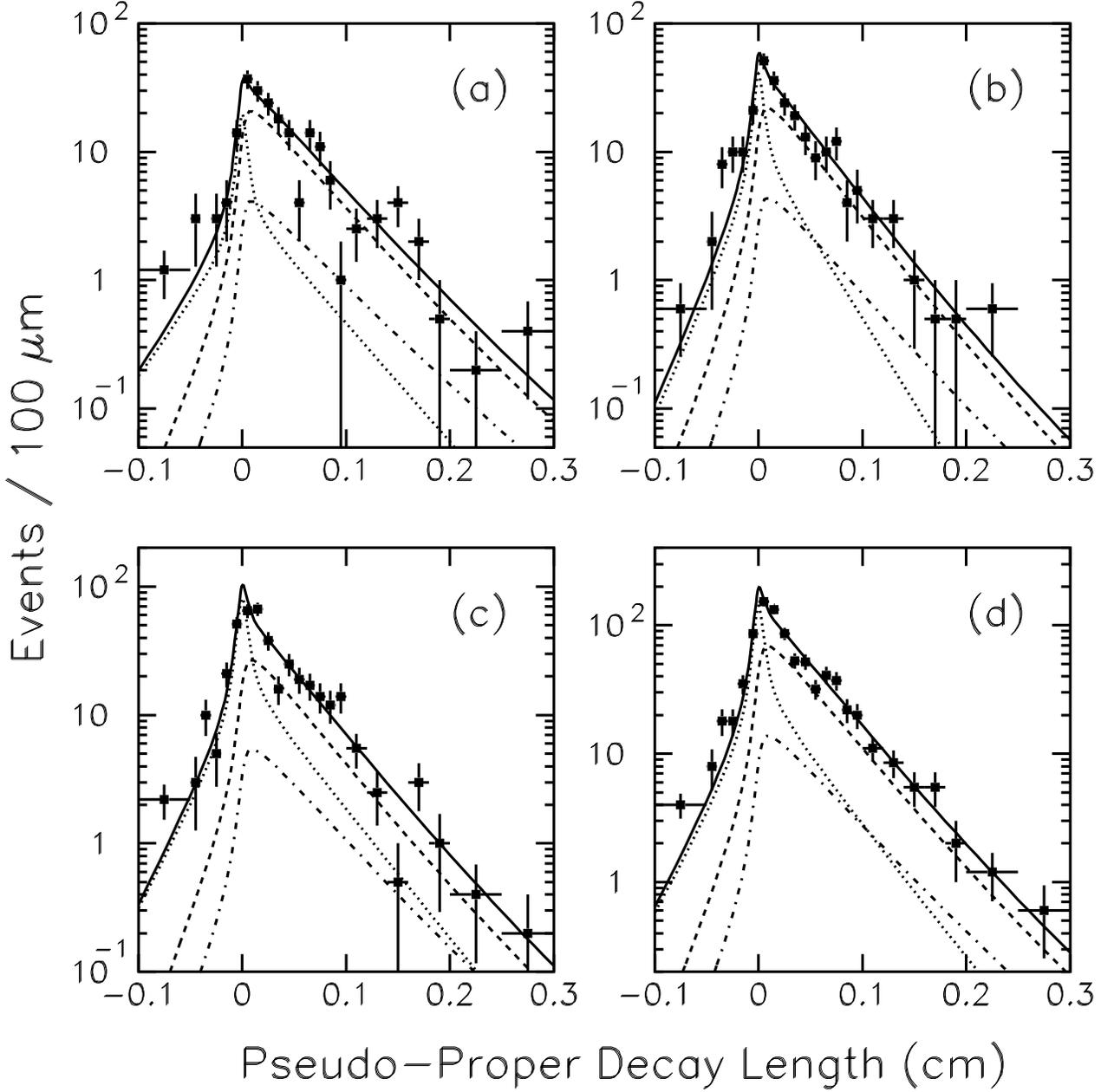}

\vspace{1in}
\caption
{Distributions of $B$ meson pseudo-proper decay lengths 
for $\ell^- D^{*+}$ signal samples (points).
Three decay modes are shown:
(a) $D^{*+} \rightarrow D^0 \pi^+$, $D^0 \rightarrow K^- \pi^+$,
(b) $D^{*+} \rightarrow D^0 \pi^+$, $D^0 \rightarrow K^- \pi^+ \pi^+ \pi^-$,
and
(c) $D^{*+} \rightarrow D^0 \pi^+$, $D^0 \rightarrow K^- \pi^+ \pi^0$.
The three modes are combined in (d). 
Also shown are the results of lifetime fits:
the ${\overline B}$$^0$ component (dashed curve),
the $B^-$ component (dot-dashed curve),
the background component (dotted curve), and
the sum of all components (solid curve).
}
\label {fig:sig}
\end {figure}

The pseudo-proper decay length distributions of the background samples
are shown in Fig.~\ref{fig:bg}, 
together with fit results.
The background parameter values and the resolution scale $s$ 
determined from the fit are listed in
Table~\ref{tb:shapes}.
The 
corresponding
decay length distributions of the signal samples
are shown in Fig.~\ref{fig:sig}.
We find the lifetimes to be
$c\tau ({\overline B}$$^0 ) = 470 \pm 44, 407 \pm 40$ and
$419 \pm 39$ $\mu$m for the three $D^0$ decay modes.
The quoted uncertainties are statistical only.
When a combined fit to the three 
modes
is made, we find
$c\tau( {\overline B}$$^0 ) = 433 \pm 24$ $\mu$m.
These results are consistent with the world average
value of $468 \pm 12$~$\mu$m~\cite{PDG98}.

\section { $B^0 {\overline B}$$^0$ mixing measurement }

The probability that a ${\overline B}$$^0$ meson at $t=0$ 
decays as ${\overline B}$$^0$ (unmixed)
or as $B^0$ (mixed) 
at a proper time $t$ is given by
\begin{eqnarray*}
P_{\rm UNM} (t) & = &
	\frac{1} {2 \tau} \exp 
\left( - \frac{t}{\tau} \right)
		( 1 + \cos \Delta m_d \, t), 
\nonumber \\
P_{\rm MIX} (t) & = &
	\frac{1} {2 \tau} \exp 
\left( - \frac{t}{\tau} \right)
		( 1 - \cos \Delta m_d \, t),
\end{eqnarray*}
where $\tau$ is the 
${\overline B}$$^0$ meson 
lifetime,
and we have ignored $CP$ violation and 
the width difference $\Delta \Gamma$ 
between the two mass eigenstates
of the $B^0 \overline{B}$$^0$ system. 
We determine the mixing parameter $\Delta m_d$ by a simultaneous fit
to the decay length distributions of unmixed and mixed decay events.


We have reconstructed the ${\overline B}$$^0$ meson decay 
${\overline B}$$^0 \rightarrow \ell^- {\bar \nu} D^{*+} X$. 
The charge of the lepton identifies 
the flavor of the $B^0$ meson at its decay.
In order to infer 
the 
${\overline B}$$^0$ meson 
flavor at its production, 
we use the second lepton candidate, which
is presumed to originate 
from the other $B$ hadron in the event.
When the other $B$ hadron, 
containing 
the ${\bar b}$ quark,
decays semileptonically, 
it 
produces    
a positively charged lepton $\ell^+$.
If the ${\overline B}$$^0$ meson,
reconstructed in the $\ell^- {\bar \nu} D^{*+} X$ decay mode,
decayed
in an unmixed state, 
the two leptons in the event 
would have the opposite charge.
Similarly, if 
the ${\overline B}$$^0$ meson decayed in
a mixed state, the leptons would have the same charge.
Therefore, 
in the ideal case, 
the opposite-sign (OS) events identify the unmixed decays
of the ${\overline B}$$^0$ meson,
while the same-sign (SS) events identify the mixed 
decays. 
However, the second lepton can originate from the
sequential decay of $B$ hadrons, ${\bar b} \rightarrow {\bar c} 
\rightarrow \ell^- X$, 
or from a mixed decay of the neutral $B$ mesons,
${\bar b} \rightarrow 
B^0$ $(B_s^0)
\rightarrow 
{\overline B}$$^0$  $({\overline B}$$^0_s)
\rightarrow \ell^- X$.
The 
lepton candidate could 
also 
be a misidentified hadron.
In these cases the second lepton candidate
will not identify the production flavor correctly.
In order to account for  these 
possibilities,
we introduce 
the probability 
of flavor misidentification
and denote it by $W$.

As mentioned above, we classify events depending on 
the sign (OS or SS) of the two lepton candidates in an event.
A finite flavor misidentification probability $W$
results in 
moving unmixed decay 
to the same-sign sample
and mixed decays to the opposite-sign sample.
Thus, we obtain the following probability distributions for the
opposite-sign and same-sign events:
\begin{eqnarray*}
P^{\rm OS} (t) & = & (1-W) P_{\rm UNM} (t)  + W P_{\rm MIX} (t) 
=
	\frac{1} {2 \tau} \exp 
\left( - \frac{t}{\tau} \right)
		[ 1 + (1-2W) \cos \Delta m_d \, t \,],
\nonumber \\
P^{\rm SS} (t) & = & (1-W) P_{\rm MIX} (t)  + W P_{\rm UNX} (t) 
=
	\frac{1} {2 \tau} \exp 
\left( - \frac{t}{\tau} \right)
		[ 1 - (1-2W) \cos \Delta m_d \, t \, ] .
\end{eqnarray*}
From these 
expressions it is evident  
that the flavor misidentification  probability
does not affect the oscillation frequency, 
although it does reduce 
its amplitude 
by a factor $1-2W$.
We  determine the two quantities that appear  
in the above expression,
$\Delta m_d$ 
and 
$W$,
simultaneously 
from the data sample
by examining the 
decay length distributions.

\subsection { $\Delta m_d$ fit}
We use the maximum likelihood method 
to extract the oscillation frequency 
$\Delta m_d$.
The likelihood is given by
${\cal L} =  \prod_i \, {\cal F} (x_i)$,
where $x_i$ is the pseudo-proper decay length measured for 
event $i$, and 
the product is taken over events in the signal sample.
The likelihood function ${\cal F} (x)$
is expressed as follows,
depending on the sign (OS or SS) of an event:
\[
{\cal F} (x)  = \left \{  \begin{array} {ll}

(1-f_{\rm BG}) \, {\cal F}_{\rm SIG} ^{\rm OS} (x)
 + f_{\rm BG} (1 - f_{\rm SS} ) \, {\cal F}_{\rm BG} (x) 
			& \;\; {\rm if} \;\; {\rm OS},\\
(1-f_{\rm BG}) \, {\cal F}_{\rm SIG} ^{\rm SS} (x)
 + f_{\rm BG} f_{\rm SS} \, {\cal F}_{\rm BG} (x) 
			& \;\; {\rm if} \;\; {\rm SS},
\end{array}
\right.  %
\]
where
$f_{\rm BG}$ is the fraction of combinatorial background events
in the sample.
The background function ${\cal F}_{\rm BG} (x)$
is the same as in the lifetime fit (Eq.~(\ref{eq:life})),
and 
its shape is taken to be the same for
both opposite-sign and same-sign events.
$f_{\rm SS}$ is the fraction of same-sign
events in the combinatorial background.

Each of the signal functions
consists of two components,
one for the ${\overline B}$$^0$ meson and the other for the $B^-$ meson:
\begin{eqnarray*}
{\cal F}_{\rm SIG}^{\rm OS} (x) &= &
               (1-g^-) \,[\, (1-W) \,{\cal F}_{\rm UNM}^0 (x)
			     +  W  \,{\cal F}_{\rm MIX}^0 (x)  \, ]
	+	    g^-      (1-W) \,{\cal F} ^- (x),
\nonumber \\
{\cal F}_{\rm SIG}^{\rm SS} (x) &= &
               (1-g^-) \,[\, (1-W) \,{\cal F}_{\rm MIX}^0 (x)
			   +   W   \,{\cal F}_{\rm UNM}^0 (x)
			                                       \, ]
	+	    g^-      W     \,{\cal F} ^- (x),
\end{eqnarray*}
where $g^-$ is the fraction of $B^-$ meson decays among the signal,
and $W$ is the flavor misidentification probability of the second
lepton.
The ${\overline B}$$^0$ component of the opposite-sign function
${\cal F}_{\rm SIG}^{\rm OS} (x)$ contains 
two terms: the first term
represents correctly tagged unmixed decays (probability $1-W$),
while the second term
represents incorrectly tagged mixed decays (probability $W$).
Similarly, 
the ${\overline B}$$^0$ component of 
the same-sign function ${\cal F}_{\rm SIG}^{\rm SS} (x)$
consists of
correctly tagged mixed decays and
incorrectly tagged unmixed decays.
Since the $B^-$ meson does not mix,
it appears in the opposite-sign function 
when the production flavor is tagged correctly,
and 
in the same-sign function  when tagged incorrectly.
The $B^-$ function 
${\cal F} ^- (x)$
is a smeared exponential decay function
and is the same as 
in the lifetime fit (Eq.~(\ref{eq:life_signal})).
The ${\overline B}$$^0$ functions
have an additional factor 
for the mixing  
and are given by
\[
{\cal F}_{\rm UNM,MIX}^0 (x) = 
\int dK \, D(K) 
\left \{
\theta(x)
\frac { K} { 2 c \tau \langle K \rangle } 
\exp \left( - \frac {K x} { c \tau \langle K \rangle }  \right)
\left [ 1 \pm \cos 
\left( \frac {\Delta m_d } {c} \frac{K} { \langle K \rangle } x \right )
\right ]
\otimes G(x)
\right \}
,
\]
where the sign $+$ ($-$) before the cosine 
corresponds to 
the 
unmixed (mixed) decay
function.
The background and $B^-$ functions are normalized 
so as to give unity when integrated over $x$.
The $B^0$ functions give unity 
when integrated over $x$ 
and summed over the  
two decay possibilities, unmixed and mixed.  
The free parameters in the fit are the oscillation frequency $\Delta m_d$
and the flavor misidentification probability $W$.
We fix the shape of the background function
(parameters $f_{\pm}$ and $\lambda_{\pm}$)
and the resolution scale factor as determined 
from the background sample.
The lifetime of the $B^0$ meson is also fixed to the value
determined earlier in the signal sample.
This procedure has been found~\cite{Bs_limits_aleph} to improve slightly
the sensitivity
in $\Delta m_d$ determination.
It is confirmed with our study using Monte Carlo events.
The background fraction $f_{\rm BG}$ in the sample
and the same-sign fraction $f_{\rm SS}$
of the background are 
fit  
parameters, but they are
constrained by 
adding a $\chi^2$ term  
to the negative log-likelihood $- \ell = - \ln {\cal L}$,
as in the lifetime fits.

\begin{figure}[p]
\vspace{2in}
\epsfysize=6.5in
\epsffile[ 50 144 522 648]{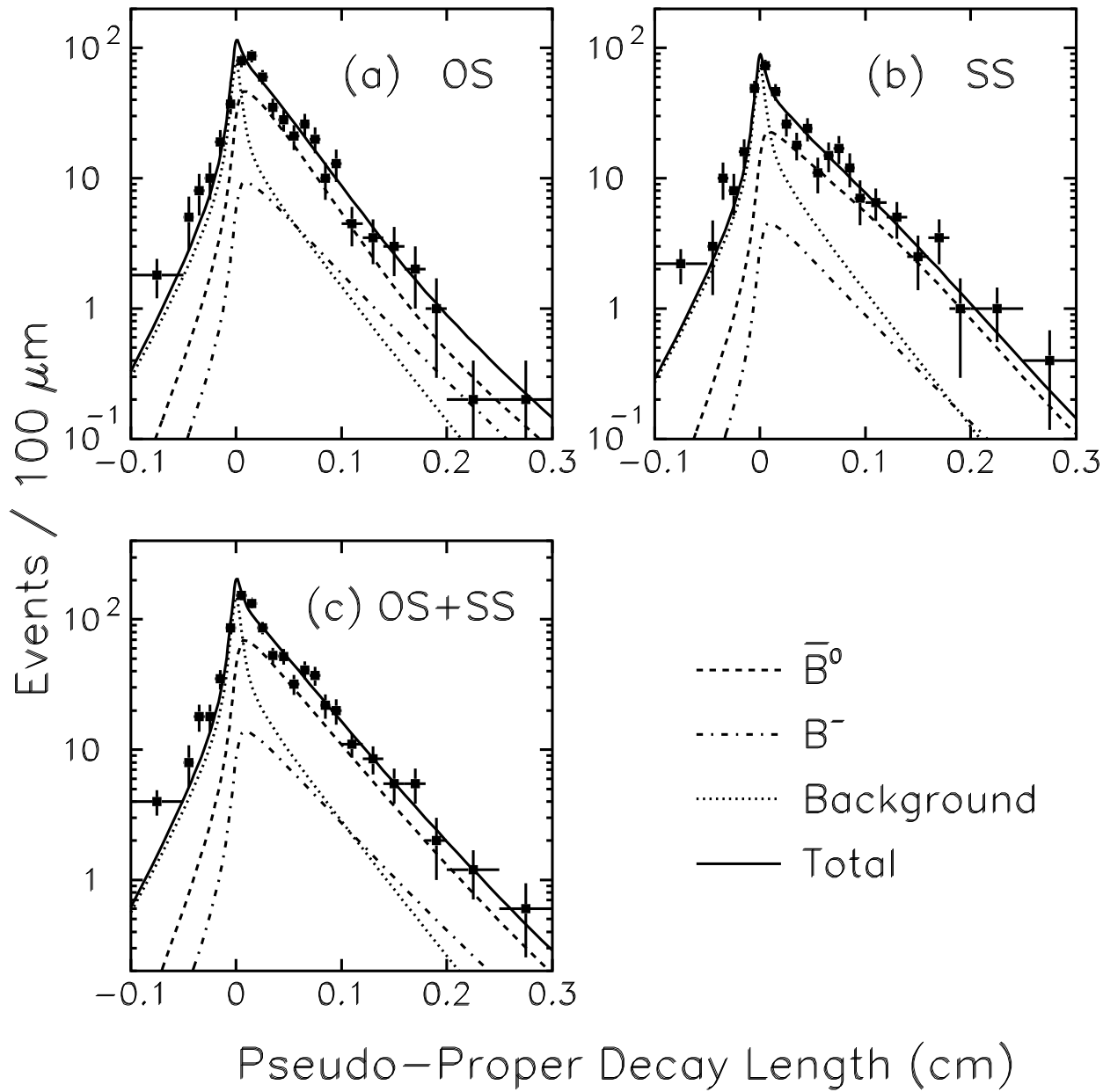} 
\vspace{1in}
\caption{ 
$B$ meson pseudo-proper decay length 
distribution (points)
estimated from the $\ell^- D^{*+}$ candidates
for (a) opposite-sign events,
and (b) same-sign events,
and (c) the sum of the two.
Curves show the result of the $\Delta m_d$ fit:
the ${\overline B}$$^0$ component (dashed curve),
the $B^-$ component (dot-dashed curve),
the background component (dotted curve), and
the sum of all components (solid curve).
}
\label {fig:decay_lengths}
\end {figure}

\begin{figure}[p]
\vspace{2in}
\epsfysize=6.5in
\epsffile[ 50 144 522 648]{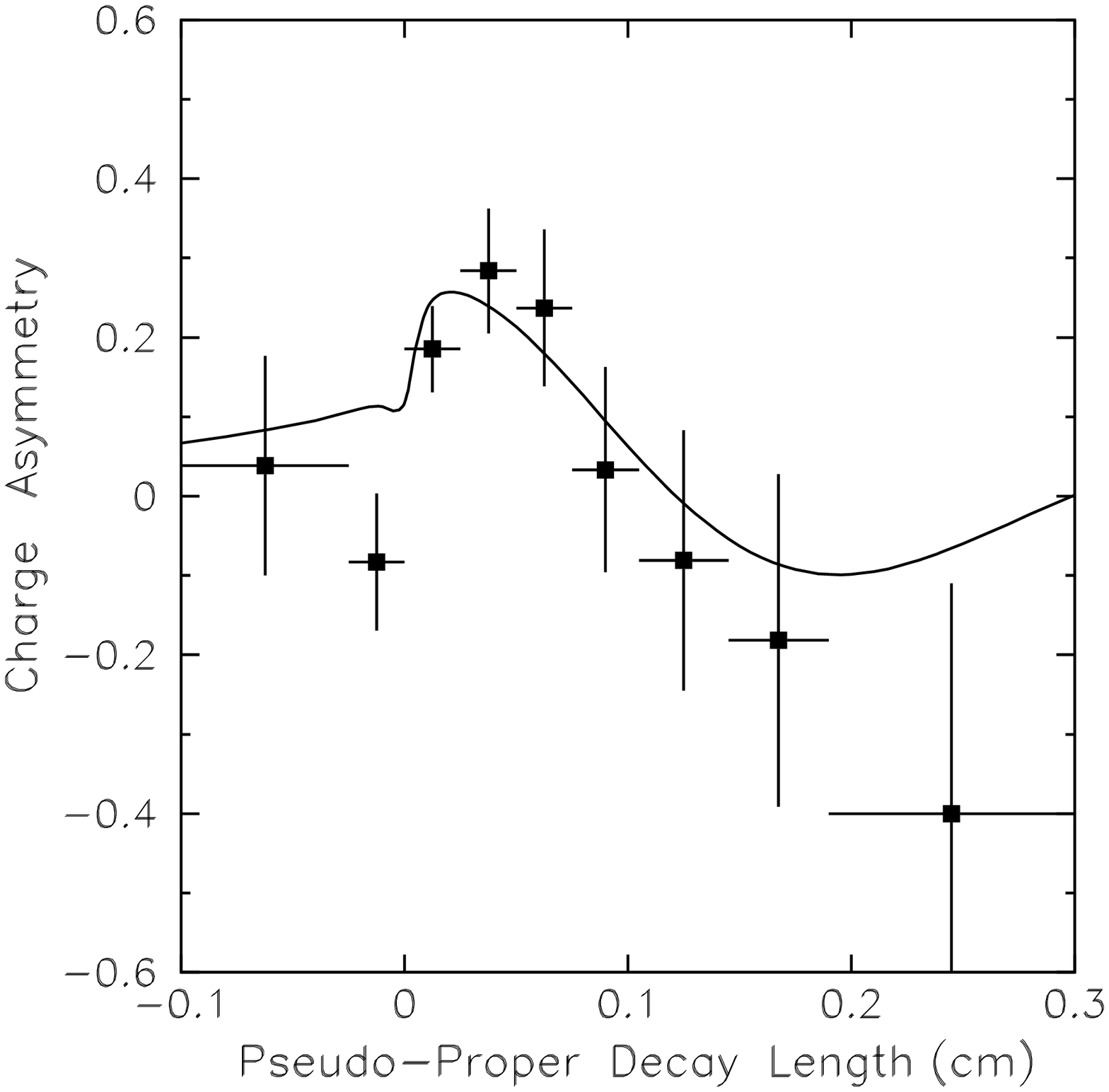} 
\vspace{1in}
\caption{ 
Charge asymmetry of the $\ell^- D^{*+}$ candidates
as a function of pseudo-proper decay length (points).
The solid curve shows the result of the $\Delta m_d$ fit.
}
\label {fig:asymmetry}
\end {figure}

The $B$ meson pseudo-proper decay length distributions 
of the $\ell^- D^{*+}$ signal sample 
are shown in Fig.~\ref{fig:decay_lengths}
for the opposite-sign and same-sign events, 
as well as 
for
the sum of the two.
The three $D^0$ decay modes are combined there. 
We find 498 opposite-sign events and 390 same-sign events.
The fraction of same-sign events in the combinatorial background,
$f_{\rm SS}$,
is estimated using events in the background sample.
The estimated same-sign 
fractions 
$f_{\rm SS}$ 
are 
summarized in Table~\ref{tb:signal2}.
The fit results are
$\Delta m_d = 
0.516  \pm 0.099 $~ps$^{-1}$
and 
$W   = 0.325 \pm 0.033$,
where 
uncertainties are statistical only.
They are summarized in Table~\ref{tb:results} along with
other 
fit parameters.
The fit results are also shown in 
Fig.~\ref{fig:decay_lengths}. 
The oscillatory 
behavior can 
be seen more directly
when the asymmetry between the opposite-sign and same-sign 
events is examined
as a function of the pseudo-proper decay length.
We define the asymmetry as
\[
A(x) = \frac 
{ N^{\rm OS}(x) - N^{\rm SS}(x) }
{ N^{\rm OS}(x) + N^{\rm SS}(x) },
\]
where 
$N^{\rm OS}(x)$ ($N^{\rm SS}(x)$)  is the number of 
opposite-sign (same-sign) events.
In the 
ideal case 
where the backgrounds, the flavor misidentification,
and the decay length smearing are absent,
the asymmetry is 
given by $A(x) = \cos ( \Delta m_d \, x /c )$.
The asymmetry distribution of the signal sample
is illustrated in Fig.~\ref{fig:asymmetry},
together with the fit result.

\subsection {Systematic uncertainties}

The sample composition is a source
of systematic uncertainty in the oscillation frequency 
measurement.
We have described it in terms of the 
parameters
$f^{**}$, $P_V$ 
and the lifetime ratio $\tau  (B^-)/ \tau ( {\overline B}$$^0 ) $.
We change 
each one of 	
the parameters  
%
to another value
while keeping the others at their nominal values,
compute the sample composition $g^-$,
and repeat the fit 
procedure for $\Delta m_d$.
We note that
the momentum correction factors ($K$ distributions)
need to be modified accordingly;
the $K$ distributions for the decay
${\overline B} \rightarrow \ell^- {\bar \nu} D^{**}$
have lower mean values 
because of additional missing particle(s), 
and changing the amount of $D^{**}$ decays  
results in changes in the $K$ distributions.
The results are summarized in Table~\ref{tb:sample}.
We interpret the observed changes as systematic uncertainties.

Other sources of systematic uncertainties 
considered in this analysis are 
summarized in Table~\ref{tb:systematics}.
Physics background processes
are studied by adding their simulated decay
length distributions to the background function.
In addition, the 
shapes of the 
decay length distributions 
of the combinatorial 
background events 
and the signal lifetime 
are subject to uncertainty because they are determined with
finite statistical precision.
They 
are changed within uncertainties, 
and 
the 
fit is repeated.
We interpret 
the 
observed changes as 
the systematic uncertainty due to this source.

Other sources of systematic uncertainties 
include our 
estimates 
of
the decay length resolution 
and of the $B$ meson momentum.
We have introduced a resolution scale factor $s$ and find a value
of about 1.2. 
We change this factor to 1.0 or 1.4 
and repeat the fit. 
We assign the observed changes as an uncertainty.
The $B^0$ meson momentum estimate ($K$ distribution) is
subject to some uncertainty too, 
because it depends on
the kinematics of $B$ meson production 
and 
of semileptonic decays.
We investigate different production and decay 
models
using the procedure described in Ref.~\cite{my_lifetimes},
and estimate the uncertainty in the $B^0$ meson momentum  
to be  
2\%,   which translates directly to the $\Delta m_d$ uncertainty.

All contributions are added in quadrature 
to give the total systematic uncertainty in 
$\Delta m_d$ of 
$^{+ \, 0.029}_{-\, 0.035}$~ps$^{-1}$  
and in the flavor misidentification probability $W$
of $^{+ \,0.006}_{- \, 0.012}$.

\section {Conclusion}

We have measured 
$B^0 {\overline B}$$^0$ mixing
using the semileptonic decay 
${\overline B}$$^0 \rightarrow \ell^- {\bar \nu} D^{*+} X$
reconstructed among ${\bar p} p$ collision events 
with two 
lepton candidates. 
The proper decay length
is estimated from reconstructed decay vertices and the 
momentum of the $\ell^- D^{*+}$ system.
A high $B^0$
purity and a relatively good momentum resolution
are achieved. The second lepton candidate in the event
is used to infer the flavor of the ${\overline B}$$^0$ meson at 
the time of its production,
with a flavor misidentification probability of
$ W =  0.325  \pm 0.033 \, ^{+ \, 0.006} _{- \, 0.012 }$.  
The frequency of the oscillation 
is measured to be
\[
\Delta m_d  =  
0.516 \pm 0.099 \,
		      ^{ + \, 0.029} _{ - \, 0.035 } 
		\;\; {\rm ps} ^{-1},
\]
where 
the first uncertainty is statistical 
and the second is systematic.
The result is consistent with
other recent measurements~\cite{new_LEP,new_LEP_common,CDF_mix}.

The method could be also applied 
in the future
to a search for 
$B^0_s {\overline B}$$^0_s$ oscillations 
with a modest value of $\Delta m_s$
by reconstructing the $D^+_s$ meson
produced in
the semileptonic decay 
${\overline B}$$^0_s \rightarrow \ell^- {\bar \nu} D_s^+ X$.

\begin{acknowledgements}
     We thank the Fermilab staff and the technical staffs of the
participating institutions for their vital contributions. This work was
supported by the U.S. Department of Energy and National Science 
Foundation; the Italian Istituto Nazionale di Fisica Nucleare; the 
Ministry of Education, Science and Culture of Japan; the Natural 
Sciences and Engineering Research Council of Canada; the National 
Science Council of the Republic of China; the A. P. Sloan Foundation; 
and the Alexander von Humboldt-Stiftung.
\end{acknowledgements}

\vspace{0.3in}
\renewcommand{\baselinestretch}{1.00}

\rm

\clearpage 

\begin {table}
\caption { Definition of signal samples, 
numbers of candidates and estimated background fractions.
}
\begin{center}
\begin {tabular} {llccrc} 
 $B$ mode & $D^0$ mode		& $D^0$ mass range & $\Delta M$ range 
& Events 
& Background fraction \\
&			& (GeV/$c^2$)      & (GeV/$c^2$)
& & \\ \hline

$\ell^- D^{*+}$ 
& $K^- \pi^+$ 
&   1.83 $-$ 1.90 & 0.144 $-$ 0.147
& 216 & $0.227 \pm 0.036$ \\

 $\ell^- D^{*+} $
& $K^- \pi^+\pi^+\pi^-$
&    1.84 $-$ 1.88 & 0.144 $-$ 0.147
& 256 	& $0.326 \pm 0.040$ \\ 

$\ell^- D^{*+} $
& $K^- \pi^+ \pi^0$
&    1.50 $-$ 1.70  & $< 0.155$
& 416 & $0.543 \pm 0.050$ \\
\end{tabular}
\end{center}
\label {tb:signal}
\end{table}

\begin {table}
\caption { Definition of background samples
and numbers of events.
}
\begin{center}
\begin {tabular} {llcccr} 
$B$ mode & $D^0$ mode & $D^0$ mass range 
& \multicolumn{2}{c} { $\Delta M$ range {(GeV/$c^2$)} }
& Events  \\ \cline{4-5}
  &    & (GeV/$c^2$)      & $D^0 \pi^+$ & $D^0 \pi^-$  & \\ \hline

$\ell^- D^{*+}$
&$K^- \pi^+$ 
&   1.83 $-$ 1.90  &  0.15 $-$ 0.19  & $<$ 0.19  &  2418 \\

 $\ell^- D^{*+}$
&$K^- \pi^+\pi^+\pi^-$
&   1.84 $-$ 1.88  &  0.15 $-$ 0.19 &   $<$ 0.19  &  5139 \\

$\ell^- D^{*+}$
&$K^- \pi^+ \pi^0$
&   1.50 $-$ 1.70  &  0.16 $-$ 0.19  &   $<$ 0.19  & 1663 \\
\end{tabular}
\end{center}
\label {tb:control}
\end{table}

\begin {table}
\caption { Background shapes obtained from 
a simultaneous fit to signal and background samples. 
}
\begin{center}
\begin {tabular} {llccccc}  
 $B$ mode & $D^0$ mode & scale $s$ & $ f_+$ & $\lambda_+$ ($\mu$m)  
			& $ f_-$ & $\lambda_-$ ($\mu$m)  
\\ \hline 

$\ell^- D^{*+}$
&$K^- \pi^+$ 
  & $1.21 \pm 0.05$ & $0.361 \pm 0.015$ & $ 474 \pm 20$
			 & $0.157 \pm 0.015$ & $ 392 \pm 49$ \\

 $\ell^- D^{*+}$
&$K^- \pi^+\pi^+\pi^-$
 & $1.17 \pm 0.03$ & $0.332 \pm 0.012$ & $ 331 \pm 11$
			  & $0.098 \pm 0.010$ & $ 230 \pm 22$\\

$\ell^- D^{*+}$
&$K^- \pi^+ \pi^0$
 & $1.21 \pm 0.04$ & $0.367 \pm 0.016$ & $ 433 \pm 22$ 
			 & $0.095 \pm 0.014$ & $ 293 \pm 33$ \\
\end{tabular}
\end{center}
\label {tb:shapes}
\end{table}

\begin {table}
\caption { 
Numbers of opposite-sign (OS) and same-sign (SS) events in the signal 
and background samples,
and the 
fraction $f_{\rm SS}$
of same-sign events in the 
combinatorial background. 
}
\begin{center}
\begin {tabular} {llrrrcrrc}  
 $B$ mode & $D^0$ mode & \multicolumn{3}{c}{Signal}
& & \multicolumn{3}{c}{Background} \\ \cline{3-5} \cline{7-9} 
 & & Sum 
& OS & SS & & OS & SS &  $ f_{\rm SS} $ 
\\ \hline 

$\ell^- D^{*+}$
&$K^- \pi^+$ 
 & 216 & 121 & 95  & 
& 1240 & 1178 & 
$0.487 \pm 0.010 $ \\

 $\ell^- D^{*+}$
&$K^- \pi^+\pi^+\pi^-$
 & 256 & 146 & 110 & 
& 2501 & 2638 & 
$0.513 \pm 0.007$ \\

$\ell^- D^{*+}$
&$K^- \pi^+ \pi^0$
 & 416 & 231 & 185 & 
& 902 & 761 & 
$0.458 \pm 0.012$ \\

$\ell^- D^{*+}$
&  Total 
 & 888 & 498 & 390 &   
 & & 
\end{tabular}
\end{center}
\label {tb:signal2}
\end{table}

\begin {table}
\caption { Results of the $\Delta m_d$ fit.
}
\begin{center}
\begin {tabular} {lcc}  
 Parameter & Input & Output 
\\ \hline 

 $\Delta m_d$ & & $0.516 \pm 0.099$ \\

 $W$ 		&    &   $0.325 \pm 0.033$ \\

 $f_{\rm BG}$ ($D^0 \rightarrow K^- \pi^+$)  
		&  $0.227 \pm 0.036$   &   $0.218 \pm 0.030$ \\
 $f_{\rm BG}$ ($D^0 \rightarrow K^- \pi^+ \pi^+ \pi^-$)  
		&  $0.326 \pm 0.040$   &   $0.355 \pm 0.032$ \\
 $f_{\rm BG}$ ($D^0 \rightarrow K^- \pi^+ \pi^0$)  
		&  $0.543 \pm 0.050$   &   $0.489 \pm 0.034$ \\


 $f_{\rm SS}$ ($D^0 \rightarrow K^- \pi^+$)  
		&  $0.487 \pm 0.010$   &   $0.488 \pm 0.010$ \\
 $f_{\rm SS}$ ($D^0 \rightarrow K^- \pi^+ \pi^+ \pi^-$)  
		&  $0.513 \pm 0.007$   &   $0.513 \pm 0.007$ \\
 $f_{\rm SS}$ ($D^0 \rightarrow K^- \pi^+ \pi^0$)  
		&  $0.458 \pm 0.012$   &   $0.465 \pm 0.011$ 
\end{tabular}
\end{center}
\label {tb:results}
\end{table}

\begin {table}
\caption{
Measurement of $\Delta m_d$ 
under various sample composition conditions.
Quoted uncertainties are statistical only.
}
\label {tb:sample}
\begin{center}
\begin {tabular} {ccccccc} 
$f^{**}$  & $P_V$ & $\tau(B^-)$ 
& \multicolumn{2}{c} {$g^-$ } 
& $\Delta m_d$  & $W$ 
\\ \cline{4-5}

& &  ${ \overline { \tau ( {\overline B}\,\!^0 ) } } $ 
& $\mu^- D^{*+}$ & $e^- D^{*+}$ &  (ps$^{-1}$) &   \\ \hline 

0.24 & 0.65 & 1.02 & 0.121 & 0.087 & $0.497 \pm 0.093$ & $0.323 \pm 0.033$  \\ 
0.36 & 0.65 & 1.02 & 0.187 & 0.138 & $0.516 \pm 0.099$ & $0.325 \pm 0.033$  \\ 
0.48 & 0.65 & 1.02 & 0.263 & 0.202 & $0.536 \pm 0.108$ & $0.327 \pm 0.033$  \\ 
\hline

0.36 & 0.26 & 1.02 & 0.090 & 0.063 & $0.488 \pm 0.090$ & $0.323 \pm 0.033$  \\ 
0.36 & 1.00 & 1.02 & 0.250 & 0.190 & $0.532 \pm 0.106$ & $0.327 \pm 0.033$  \\ 
\hline

0.36 & 0.65 & 0.97 & 0.179 & 0.132 & $0.511 \pm 0.098$ & $0.324 \pm 0.033$  \\ 
0.36 & 0.65 & 1.07 & 0.194 & 0.144 & $0.520 \pm 0.101$ & $0.325 \pm 0.033$  \\
\end{tabular}
\end{center}
\end{table}

\begin{table}  
\caption{
A summary of systematic uncertainties in $\Delta m_d$ measurement.}
\begin{center}
\begin{tabular}{lcc} 
Source            & \multicolumn{2}{c} 
                                        {Contribution to} \\ \cline{2-3}
	          & $\Delta m_d$ (${\rm ps}^{-1}$)  & $W$ \\ \hline
Sample composition            &            \\
\ \ \ \ $D^{**}$ fraction ($f^{**}$)  &
        $^{+0.020}_{-0.018}$ &
        $^{+0.003}_{-0.001}$ \\
\ \ \ \ $D^{**}$ composition ($P_V$)  &
        $^{+0.017}_{-0.027}$ &
        $^{+0.002}_{-0.001}$ \\
\ \ \ \ Lifetime ratio $\tau(B^-) / \tau( {\overline B}$$^0 )$ &
        $\pm$ 0.005           &
        $\pm$ 0.001             \\
\ \ \ \ Physics background & 
	$^{+0.003}_{-0.000}$  &
	$^{+0.000}_{-0.010}$ \\


Background shape, $\overline{B}$$^0$ lifetime  &
	$\pm$ 0.001 &
	$^{+0.003}_{-0.007}$ \\
Decay length resolution  & 
	$\pm$ 0.005 &
	$^{+0.003}_{-0.002}$ \\

$B$ meson momentum estimate &
	$\pm$ 0.011&
	-  \\ \hline

Total           & 
	$^{+0.029}_{-0.035}$ &	
	$^{+0.006}_{-0.012}$ 	\\ 
\end{tabular}
\end{center}
\label{tb:systematics}
\end{table}

\clearpage

\clearpage

\clearpage

\clearpage

\clearpage

\clearpage

\clearpage



\begin{thebibliography}{99}

\bibitem{Argus_mix}
ARGUS Collaboration, 
	H.~Albrecht {\em et al.}, Phys. Lett. B {\bf 192}, 245 (1987);
CLEO Collaboration, 
	M.~Artuso   {\em et al.}, Phys. Rev. Lett. {\bf 62}, 2233 (1989).


\bibitem{Lim}
T.~Inami and C.~S.~Lim, Prog. Theor. Phys. {\bf 65}, 297 (1981);
{\bf 65}, 1772(E) (1981).


\bibitem{CKM}
M.~Kobayashi and T.~Maskawa, Prog. Theor. Phys. {\bf 49}, 652 (1973).


\bibitem{chi_d}
ARGUS Collaboration, 
	H.~Albrecht {\em et al.}, Z. Phys. C {\bf 55}, 357 (1992);
CLEO Collaboration, 
	J.~Bartelt  {\em et al.}, Phys. Rev. Lett. {\bf 71}, 1680 (1993);
ARGUS Collaboration, 
	H.~Albrecht {\em et al.}, Phys. Lett. B {\bf 324}, 249 (1994).


\bibitem{new_LEP}
L3 Collaboration, 
	M.~Acciarri {\em et al.}, 
	Eur. Phys. J. C {\bf 5}, 195  (1998);
DELPHI Collaboration, 
	P.~Abreu {\em et al.}, 
	Z. Phys. C {\bf 76}, 579 (1997);
OPAL Collaboration, 
	K.~Ackerstaff {\em et al.}, 
	Z. Phys. C {\bf 76}, 417 (1997);
ALEPH Collaboration, 
	D.~Buskulic {\em et al.}, 
	Z. Phys. C {\bf 75}, 397 (1997);
OPAL Collaboration, 
	G.~Alexander {\em et al.}, 
	Z. Phys. C {\bf 72}, 377 (1996).

\bibitem{new_LEP_common}
OPAL Collaboration, 
	K.~Ackerstaff {\em et al.}, 
	Z. Phys. C {\bf 76}, 401 (1997).

\bibitem{CDF_mix}
CDF Collaboration, 
	F.~Abe {\em et al.}, 
	Phys. Rev. Lett. {\bf 80}, 2057 (1998);    
CDF Collaboration, 
	F.~Abe {\em et al.}, 
	Phys. Rev. D {\bf 59}, 032001 (1999);   
CDF Collaboration, 
	F.~Abe {\em et al.}, 
	Phys. Rev. D {\bf 60}, 051101 (1999);
CDF Collaboration, 
	F.~Abe {\em et al.}, 
	FERMILAB-Pub-99/019-E (hep-ex/990311) (1999),  
	to be publihed in Phys. Rev. D. 			




\bibitem{new_Bs_limits} 
ALEPH Collaboration, 
	R.~Barate {\em et al.}, 
	Eur. Phys. J. C {\bf 4}, 367 (1998);
DELPHI Collaboration, 
	W.~Adam {\em et al.}, 
	Phys. Lett. B {\bf 414}, 382 (1997);
CDF Collaboration, 
	F.~Abe {\em et al.}, 
	Phys. Rev. Lett. {\bf 82}, 3576 (1999). 





\bibitem{CDF}
CDF Collaboration, F.~Abe {\em et al.}, 
	Nucl. Instrum. Methods Phys. Res. A {\bf 271}, 387 (1988), 
	and references therein.


\bibitem{svx} 
D. Amidei {\em et al.},
	Nucl. Instrum. Methods Phys. Res. A {\bf 350}, 73 (1994);
P. Azzi {\em et al.},
	Nucl. Instrum. Methods Phys. Res. A {\bf 360}, 137 (1995).









\bibitem{lateral1} 
CDF Collaboration, F.~Abe {\em et al.}, 
	Phys. Rev. D {\bf 52}, 2624~(1995). 


\bibitem{lateral2} 
CDF Collaboration, F.~Abe {\em et al.}, Phys.~Rev.~Lett.~{\bf 68}, 2734 (1992);
CDF Collaboration, F.~Abe {\em et al.}, Phys. Rev. D {\bf 48}, 2998~(1993).



\bibitem{PDG98}
C.~Caso {\em et al.}, Eur. Phys. J. C {\bf 3},  1 (1998).


\bibitem{cdf_psi_life}
CDF Collaboration, F.~Abe {\em et al.}, 
	Phys.~Rev.~Lett.~{\bf 71}, 3421~(1993); 
	CDF Collaboration, F.~Abe {\em et al.}, 
	Phys. Rev. D {\bf 57}, 5382 (1998).   


\bibitem{my_lifetimes}
CDF Collaboration, F.~Abe {\em et al.}, 
	Phys. Rev. Lett. {\bf 76}, 4462 (1996);
CDF Collaboration, F.~Abe {\em et al.}, 
	Phys. Rev. D {\bf 58}, 092002 (1998).


\bibitem{CLEO}
CLEO Collaboration, R. Fulton {\em et al.}, 
	Phys. Rev. D {\bf 43}, 651 (1991).

\bibitem {ddst}
ALEPH Collaboration, D. Busklic {\em et al.}, 
	Phys. Lett. B {\bf 345}, 103 (1995); 
DELPHI Collaboration, P. Abreu {\em et al.}, 
	Z. Phys. C {\bf 71}, 539 (1996);  
OPAL Collaboration, R. Akers {\em et al.}, 
	Z. Phys. C {\bf 67}, 57 (1995);  
CLEO Collaboration, 
	Phys. Rev. Lett. {\bf 80}, 4127 (1998).

\bibitem {ISGW}
        N. Isgur, D. Scora, B. Grinstein, and M. Wise,
        Phys. Rev. D {\bf 39}, 799 (1989);
        D. Scora and N. Isgur, 
        Phys. Rev. D {\bf 52}, 2783 (1995).




\bibitem{PDG}
R.~M.~Barnett {\em et al.}, Phys. Rev. D {\bf 54},  1 (1996).



\bibitem{NDE}
P. Nason, S. Dawson, and R. K. Ellis, Nucl. Phys. {\bf B327}, 49 (1989);
{\bf B335}, 260(E) (1990).


\bibitem{Peterson}
C. Peterson, D. Schlatter, I. Schmitt, and P. M. Zerwas,
Phys. Rev. D {\bf 27}, 105 (1983).


\bibitem{QQ} P.~Avery, K.~Read, and G.~Trahern, 
Report No.~CSN-212, 1985 (unpublished).


\bibitem{Bs_limits_aleph} 
ALEPH Collaboration, 
	D.~Buskulic {\em et al.}, 
	Phys. Lett. B {\bf 377}, 205 (1996).



\end{thebibliography}
\end{document}